\documentclass[preprint,showpacs,floats,epsf,epsfig,nofootinbib]{revtex4}
\usepackage{epsfig}

\begin{document}
\newcommand{\sla}[1]{#1\!\!\!\slash}
\def\be{\begin{eqnarray}}
\def\en{\end{eqnarray}}
\def\non{\nonumber}
\def\la{\langle}
\def\ra{\rangle}
\def\nc{N_c^{\rm eff}}
\def\vp{\varepsilon}
\def\drho{\bar\rho}
\def\deta{\bar\eta}
\def\CP{{\it CP}~}
\def\a{{\cal A}}
\def\B{{\cal B}}
\def\c{{\cal C}}
\def\d{{\cal D}}
\def\e{{\cal E}}
\def\p{{\cal P}}
\def\t{{\cal T}}
\def\up{\uparrow}
\def\dw{\downarrow}
\def\vma{{_{V-A}}}
\def\vpa{{_{V+A}}}
\def\smp{{_{S-P}}}
\def\spp{{_{S+P}}}
\def\J{{J/\psi}}
\def\ov{\overline}
\def\Lqcd{{\Lambda_{\rm QCD}}}
\def\pr{{Phys. Rev.}~}
\def\prl{{Phys. Rev. Lett.}~}
\def\pl{{Phys. Lett.}~}
\def\np{{Nucl. Phys.}~}
\def\zp{{Z. Phys.}~}
\def\lsim{ {\ \lower-1.2pt\vbox{\hbox{\rlap{$<$}\lower5pt\vbox{\hbox{$\sim$}
}}}\ } }
\def\gsim{ {\ \lower-1.2pt\vbox{\hbox{\rlap{$>$}\lower5pt\vbox{\hbox{$\sim$}
}}}\ } }

\font\el=cmbx10 scaled \magstep2{\obeylines\hfill April, 2006}

\vskip 1.5 cm

\centerline{\large\bf Doubly Charmful Baryonic $B$ Decays}

\bigskip
 \centerline{\bf Hai-Yang Cheng, Chun-Khiang Chua and Shang-Yuu Tsai}
\medskip
\centerline{Institute of Physics, Academia Sinica}
\centerline{Taipei, Taiwan 115, Republic of China}
\medskip
\bigskip
\bigskip
\bigskip
\bigskip
\centerline{\bf Abstract}
\bigskip
\small

There are two apparent puzzles connected with the two-body and
three-body doubly charmed baryonic $B$ decays. First, earlier
calculations based on QCD sum rules or the diquark model predict
$\B(\ov B^0\to\Xi_c^+\bar\Lambda_c^-)\approx \B(\ov B^0\to\B_c\ov
N)$, while experimentally the former has a rate two orders of
magnitude larger than the latter. Second, a naive estimate of the
branching ratio ${\cal O}(10^{-9})$ for the color-suppressed
three-body decay $\ov B\to\Lambda_c^+\bar\Lambda_c^-K$, which is
highly suppressed by phase space, is too small by five to six
orders of magnitude compared to experiment. We show that the great
suppression for the $\Lambda_c^+\bar\Lambda_c^-K$ production can
be alleviated provided that there exists a narrow hidden charm
bound state with a mass near the $\Lambda_c\bar\Lambda_c$
threshold. This new state that couples strongly to the charmed
baryon pair can be searched for in $B$ decays and in $p\bar p$
collisions by studying the mass spectrum of $D^{(*)}\ov D^{(*)}$
or $\Lambda_c\bar\Lambda_c$. The doubly charmful decay $\ov
B\to\Xi_c\bar\Lambda_c$ has a configuration more favorable than
the singly charmful one such as $\ov B^0\to\Lambda_c\bar p$ since
no hard gluon is needed to produce the energetic
$\Xi_c\bar\Lambda_c$ pair in the former decay, while two hard
gluons are needed for the latter process. Assuming that a soft
$q\bar q$ quark pair is produced through the $\sigma$ and $\pi$
meson exchanges in the configuration for $\ov B\to
\Xi_c\bar\Lambda_c$, it is found that its branching ratio is of
order $10^{-3}$, in agreement with experiment.


\maketitle

\pagebreak

 \section{Introduction}
Recently Belle has observed for the first time two-body and
three-body doubly charmed baryonic $B$ decays in which two charmed
baryons are produced in the final state \cite{3body,2body}. The
measured branching ratios are
 \be
 \B(B^-\to
 \Lambda_c^+\bar\Lambda_c^-K^-) &=& (6.5^{+1.0}_{-0.9}\pm0.8\pm3.4)\times
 10^{-4}, \non \\
 \B(B^0\to \Lambda_c^+\bar\Lambda_c^-K^0) &=& (7.9^{+2.9}_{-2.3}\pm1.2\pm4.2)\times
 10^{-4},
 \en
 for three-body decays and
 \be
 \B(B^-\to\Xi_c^0\bar\Lambda_c^-)\B(\Xi_c^0\to
 \Xi^-\pi^+) &=& (4.8^{+1.0}_{-0.9}\pm1.1\pm1.2)\times 10^{-5} ,
 \non \\
 \B(\ov B^0\to\Xi_c^+\bar\Lambda_c^-)\B(\Xi_c^+\to
 \Xi^-\pi^+\pi^+) &=& (9.3^{+3.7}_{-2.8}\pm1.9\pm2.4)\times 10^{-5}
 \en
for two-body decays. Taking the theoretical estimates (see e.g.
Table III of \cite{CT93}), $\B(\Xi_c^0\to \Xi^-\pi^+)\approx
1.3\%$ and $\B(\Xi_c^+\to \Xi^0\pi^+)\approx 3.9\%$ together with
the experimental measurement $\B(\Xi_c^+\to
\Xi^0\pi^+)/\B(\Xi_c^+\to
 \Xi^-\pi^+\pi^+)=0.55\pm0.16$ \cite{PDG}, it follows
 that
  \be
 \B(B^-\to\Xi_c^0\bar\Lambda_c^-)\approx 4.8\times 10^{-3}, \qquad
 \B(\ov B^0\to\Xi_c^+\bar\Lambda_c^-)\approx 1.2\times 10^{-3}.
 \en
Therefore, the two-body doubly charmed baryonic $B$ decay $B\to
\B_c\bar \B'_c$ has a branching ratio of order $10^{-3}$, to be
compared with~\cite{Lambdacp,Lambdacppi}
 \be
 \B(\ov B^0\to\Lambda_c^+\bar p) &=&
 (2.19^{+0.56}_{-0.49}\pm0.32\pm0.57)\times 10^{-5}, \non \\
 \B(B^-\to\Sigma_c(2455)^0\bar p) &=&
 (3.67^{+0.74}_{-0.66}\pm0.36\pm0.95)\times
 10^{-5},
 \en
for singly charmed baryonic $B$ decays
and~\cite{BaBarpp,Belle2body}
 \be
 && \B(B^0\to p\bar p)<2.7\times 10^{-7}, \qquad
 \B(B^0\to \Lambda\bar\Lambda )<6.9\times 10^{-7},
 \non \\
 && \B(B^-\to\Lambda\bar p)<4.9\times 10^{-7},
 \en
for charmless baryonic $B$ decays. Therefore, we have the pattern
 \be \label{eq:2bodypattern}
 \B_c\bar \B'_c~(\sim 10^{-3})\gg \B_c\bar\B~(\sim 10^{-5})\gg
 \B_1\bar \B_2~(\sim 10^{-7}),
 \en
for two-body baryonic $B$ decays.

Using $\B(\ov B^0\to\Lambda_c^+\bar p)$ as a benchmark, one will
expect a branching ratio of order $10^{-7}$ for the charmless
decay $B\to \B_1\bar \B_2$ after replacing the quark mixing angle
$V_{cb}$ by $V_{ub}$, provided that the dynamical suppression for
the latter is neglected. However, since the doubly charmed
baryonic decay mode $\Xi_c\bar\Lambda_c$ proceeds via $b\to cs\bar
c$, while $\Lambda_c\bar p$ via a $b\to cd\bar u$ quark
transition, the CKM mixing angles for them are the same in
magnitude but opposite in sign. One may wonder why the $\B_c\bar
\B'_c$ mode has a rate two orders of magnitude larger than
$\B_c\bar\B$. Indeed, earlier calculations based on QCD sum rules
\cite{Chernyak} or the diquark model \cite{Ball} all predict that
$\B(B\to \Xi_c\bar\Lambda_c)\approx \B(\ov B\to\B_c\ov N)$, which
is in violent disagreement with experiment. This implies that some
important dynamical suppression effect for the $\B_c\ov N$
production with respect to $\Xi_c\bar\Lambda_c$ is missing in
previous studies.

As for the three-body decay $B\to \Lambda_c\bar\Lambda_c K$, its
branching ratio is estimated to be of order $10^{-9}$, which is
extremely small due to the tiny phase space available for this
decay and the color-suppression effect. The puzzle is that why the
measured rate is much larger than the naive expectation?

A crucial ingredient for understanding the baryonic $B$ decays is
the threshold or low mass enhancement behavior of the baryon-pair
invariant mass in the spectrum for $B\to \B_1\bar \B_2M$: It
sharply peaks at very low values. That is, the $B$ meson is
preferred to decay into a baryon-antibaryon pair with low
invariant mass accompanied by a fast recoiled meson. Therefore,
some three-body final states have rates larger than their two-body
counterparts, e.g. $p\bar p K^\pm\gg p\bar p,~\Lambda\bar
p\pi^\pm\gg\Lambda\bar p,~\Sigma_c\bar p \pi^\pm\gg\Sigma_c\bar
p$.\footnote{The three-body decay is usually referred to the
nonresonant one. The relation $\Lambda_c\bar
p\pi^-\gg\Lambda_c\bar p$ is trivial as the former arises mostly
from resonant contributions \cite{Lambdacppi}.}
This phenomenon can be understood in terms of the threshold
effect, namely, the invariant mass of the dibaryon is preferred to
be close to the threshold. The configuration of the two-body decay
$B\to\B_1\ov \B_2$ is not favorable since its invariant mass is
$m_B$. In $B\to \B_1\ov\B_2 M$ decays, the effective mass of the
baryon pair is reduced as the emitted meson can carry away a large
amount of energies. The two-body decay pattern
(\ref{eq:2bodypattern}) also follows from the low-mass enhancement
effect: The energy release is least for the $B$ decay into two
charmed baryons and becomes very large when the final-state
baryons are charmless.

Although the gross feature of the baryonic $B$ decays can be
qualitatively comprehended in terms of the near threshold effect,
how to quantitatively evaluate their absolute decay rates and how
to realize the low mass enhancement effect require detailed
dynamical studies. In the present work we will focus on the doubly
charmful baryonic $B$ decays, namely, $\ov B\to\Xi_c\bar\Lambda_c$
and $\ov B\to\Lambda_c\bar\Lambda_cK$ in Secs. II and III,
respectively, aiming to resolve the aforementioned two puzzles
connected with them. Sec. IV gives the conclusion. The evaluation
of the delta functions occurring in the phase space integral is
discussed in the Appendix.

\begin{figure}
\centerline{\epsfig{file=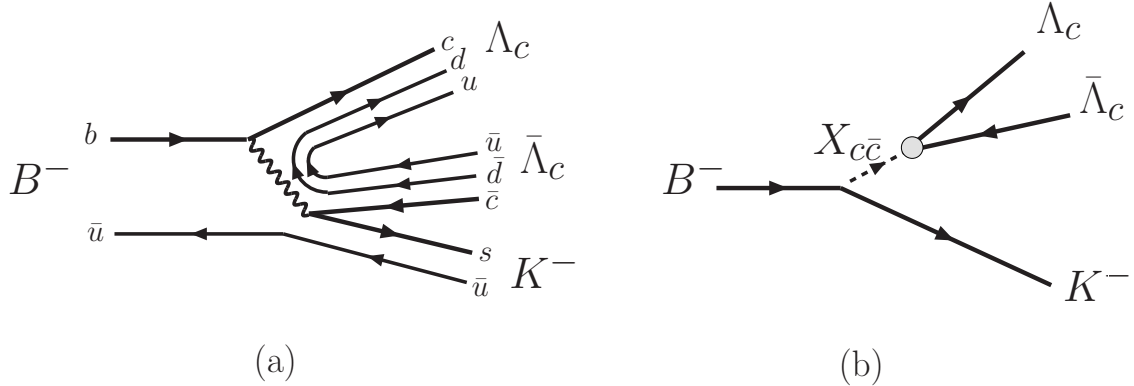,width=6in}}
\caption{\label{fig:B2LcLcK} $B^-\to\Lambda_c^+\bar\Lambda_c^-K^-$
as proceeding through (a) the internal $W$-emission diagram, and (b)
the dominant charmonium-like resonance $X_{c\bar c}$. The blob in
(b) shows where the strong decays take place.}
\end{figure}

\section{Three-body decays}
We consider the
decay $B^-\to\Lambda_c^+\bar\Lambda_c^-K^-$, which proceeds
through the internal $W$-emission diagram
Fig.~\ref{fig:B2LcLcK}(a). It turns out this diagram is
factorizable. In the weak Hamiltonian approach, the factorizable
amplitude reads
 \be
 A(B^-\to\Lambda_c^+\bar\Lambda_c^-K^-)=
 {G_F\over\sqrt{2}}V_{cs}V_{cb}^*\, a_2\la
 \Lambda_c^+\bar\Lambda_c^-|(\bar cc)|0\ra\la K^-|(\bar
 s b)|B^-\ra,
 \en
where $(\bar q_1q_2)\equiv \bar q_1\gamma_\mu(1-\gamma_5)q_2$, and
the effective Wilson coefficient $a_2$ indicates that this decay
is color suppressed. The matrix elements can be parametrized as
 \be \label{eq:baryonFF}
 \la \Lambda_c(p_1)\bar\Lambda_c(p_2)|(\bar cc)|0\ra &=&
 \bar u_{\Lambda_c}(p_1)\left[f_1(q^2)\gamma_\mu+i{f_2(q^2)\over
 2m_{\Lambda_c}}\sigma_{\mu\nu}q^\nu-\left(g_1(q^2)\gamma_\mu+{g_3(q^2)\over
 2m_{\Lambda_c}}q_\mu
 \right)\gamma_5\right]v_{\bar\Lambda_c}(p_2), \non  \\
  \la K^-(p_K)|(\bar sb)|B^-(p_B)\ra &=& F_1^{BK}
 (q^2)(p_B+p_K)_\mu+\left(F_0^{BK}
 (q^2)-F_1^{BK}(q^2)\right){m_B^2-m_K^2\over q^2}q_\mu,
 \en
with $q=p_B-p_K=p_1+p_2$. In terms of the form factors, the decay
amplitude has the expression
 \be
 A(B^-\to\Lambda_c^+\bar\Lambda_c^-K^-)={G_F\over\sqrt{2}}V_{cs}V_{cb}^*
\,a_2 \bar u_{\Lambda_c}[a
p\!\!\!/_K+b-(cp\!\!\!/_K+d)\gamma_5]v_{\bar\Lambda_c},
 \en
with
 \be
 a&=& 2F_1^{BK}(q^2)\left[f_1(q^2)+f_2(q^2)\right], \non \\
 b&=& 2F_1^{BK}(q^2)f_2(q^2)(p_2-p_1)\cdot p_K/(2m_{\Lambda_c}),
 \non \\
 c &=& 2F_1^{BK}(q^2)g_1(q^2), \non \\
 d &=& 2m_{\Lambda_c}g_1(q^2)\left[F_1^{BK}(q^2)+
 (F^{BK}_0(q^2)-F^{BK}_1(q^2)){m_B^2-m_K^2\over
 q^2}\right]  \non \\
 &&+g_3(q^2)F_0^{BK}(q^2)(m_B^2-m_K^2)/(2m_{\Lambda_c}).
 \label{eq:abcd}
 \en
There are numerous estimates of the $B\to K$ transition form
factors. We will follow \cite{CCH} where the form factors are
evaluated using the relativistic covariant light-front quark model.

Due to the heavy mass of $\Lambda_c$, the phase space of the
$\Lambda^+_c\bar\Lambda^-_cK^-$ decay is about a hundred times
smaller than, say, that of the $\Lambda\bar
p\pi^+$~\cite{Abe:2004wf}. The $\Lambda^+_c\bar\Lambda^-_c$ form
factors, if any, can therefore be taken as constants whose values
are determined at the threshold over the phase space. To achieve a
rate that is at least comparable with that of the $\Lambda\bar
p\pi^+$ whose branching ratio is of order $3\times 10^{-6}$
\cite{Abe:2004wf}, one would need the $\Lambda^+_c\bar\Lambda^-_c$
form factors to be more than one hundred times larger than those
of the $\Lambda\bar p\pi^+$ near the $\Lambda\bar p$ threshold,
which is quite unlikely since the $\Lambda\bar p$ form factors
have their maximum values already about $\mathcal{O}(1)$. Besides,
$\mathcal{O}(10^2)$ form factors would just give a rate of
$\Lambda^+_c\bar\Lambda^-_cK^-$ comparable to
$\sim\mathcal{O}(10^{-6})$, not to mention the remaining factor of
$\mathcal{O}(10^2)$ difference between the rates of $\Lambda\bar
p\pi^+$ and of $\Lambda^+_c\bar\Lambda^-_cK^-$. Therefore we
conclude that the suppression from the
$\Lambda_c^+\bar\Lambda_c^-K^-$ phase space is so strong that
$\Lambda_c^+\bar\Lambda_c^-$ pair is unlikely to be produced
dominantly through the direct three-body decay processes. The
great suppression, however, seems to hint strongly that a $c\bar
c$-content resonance with the
width comparable to the nearby resonances like $\psi(4415)$ 
could be located around the threshold of
$\Lambda_c^+\bar\Lambda_c^-~(\sim4.6~\mathrm{GeV})$, and the whole
process takes place dominantly via the charmonium-like resonance
as shown in Fig.~\ref{fig:B2LcLcK}(b).

Let us assume the resonance, $X_{c\bar c}$, exists with a mass
$m_{X_{c\bar c}}\gtrsim2m_{\Lambda_c}$ and a width
$\Gamma_{X_{c\bar c}}$. Let us further assume that this resonance
is a spin-1 particle with $J^P=1^-$ or $1^+$, as inspired from the
observation that all the charmonia near the
$\Lambda_c^+\bar\Lambda_c^-$ threshold are spin-1 particles. The
decay amplitude then reads
\be A_{X_{c\bar c}}(B^-\to\Lambda_c^+\bar\Lambda_c^-K^-)&=&
 {G_F\over\sqrt{2}}V_{cs}V_{cb}^*\, a_2\la K^-|\bar
 s\gamma_\mu(1-\gamma_5) b|B^-\ra\non\\
 &&\times  m_{X_{c\bar c}}f_{X_{c\bar c}}\left(\frac{-g^{\mu\nu}+\frac{q^\mu
q^\nu}{m_{X_{c\bar c}}^2}}{q^2-m_{X_{c\bar c}}^2+i\,m_{X_{c\bar
c}}\Gamma_{X_{c\bar c}}}\right)i\,\bar u_{\Lambda_c}(p_1)M_\nu
v_{\bar\Lambda_c}(p_2)\,,
\label{eq:decay amp}%
\en
where
\begin{equation}
M_\nu=h^{\Lambda_c\bar\Lambda_cV}_1\gamma_\nu+{ih^{\Lambda_c\bar\Lambda_cV}_2\over
 2m_{\Lambda_c}}\,\sigma_{\nu\rho}\,q^\rho\,,
\label{eq:vLcLc}
\end{equation}
when $X_{c\bar c}$ is a vector~($X_{c\bar c}=V$), and
\begin{equation}
M_\nu=\left(h^{\Lambda_c\bar\Lambda_cA}_1\gamma_\nu+{h^{\Lambda_c\bar\Lambda_cA}_2\over
 2m_{\Lambda_c}}\,q_\nu\right)\gamma_5\,,
\label{eq:aLcLc}
\end{equation}
when $X_{c\bar c}$ is an axial-vector~($X_{c\bar c}=A$) particle.
$f_{X_{c\bar c}}$ is the decay constant for $X_{c\bar c}$, and
$h^{\Lambda_c\bar\Lambda_cV}_{1,2}$ and
$h^{\Lambda_c\bar\Lambda_cA}_{1,2}$ represent the {\it
dimensionless} $\Lambda_c\bar\Lambda_cX_{c\bar c}$ strong
couplings. Since the allowed phase space is very small, the strong
couplings can effectively be treated as constants within this
region. The decay constant $f_{X_{c\bar c}}$ comes from the
factorization of the amplitude of $B^-\to X_{c\bar c}K^-$,
followed by the strong decay $X_{c\bar c}\to
\Lambda_c^+\bar\Lambda_c^-$. Since the chirality-flipping baryon
vector form factor $f_2(q^2)$ is in general suppressed by two more
powers of the dibaryon invariant mass $q^2$ than $f_1(q^2)$ and
since $q^2\sim4\,m^2_{\Lambda_c}$ is large in
$B\to\Lambda_c\bar\Lambda_cK$ decays, we expect the contributions
from $X_{c\bar c}$ coupled to $\Lambda_c\bar\Lambda_c$ through
$h^{\Lambda_c\bar\Lambda_cV}_2$ be small and hence can be
neglected in our calculation. The $h^{\Lambda_c\bar\Lambda_cA}_2$
term in the axial-vector decay amplitude can also be dropped since
$q_\nu(-g^{\mu\nu}+q^\mu q^\nu/m_A^2)\sim0$ due to the fact that
$m^2_A\sim q^2$ within the phase space. The decay amplitude
$A_{X_{c\bar c}}$ then becomes
\be
A_{V(A)}(B^-\to\Lambda_c^+\bar\Lambda_c^-K^-)={G_F\over\sqrt{2}}V_{cs}V_{cb}^*
\,a_2\,m_{V(A)}f_{V(A)} \bar u_{\Lambda_c}\left[\mathcal{M}_{V(A)}
p\!\!\!/_K\left(\gamma_5\right)+\mathcal{M}_{S(P)}\left(\gamma_5\right)\right]v_{\bar\Lambda_c},
\en
with
\be
\mathcal{M}_V &=& 2F_1^{BK}(q^2)\left(\frac{-h^{\Lambda_c\bar\Lambda_cV}_1}{q^2-m_V^2+i\,m_V\Gamma_V}\right), \non \\
\mathcal{M}_S&=& 0,
 \non \\
\mathcal{M}_A &=& 2F_1^{BK}(q^2)\left(\frac{-h^{\Lambda_c\bar\Lambda_cA}_1}{q^2-m_A^2+i\,m_A\Gamma_A}\right), \non \\
\mathcal{M}_P &=& \left[F_1^{BK}(q^2){q^2-(m_B^2-m_K^2)\over
 q^2}+
 F^{BK}_0(q^2)(m_B^2-m_K^2)\left({1\over
 q^2}-\frac{1}{m_A^2}\right)\right]
 \non \\
&&
\times2m_{\Lambda_c}\left(\frac{-h^{\Lambda_c\bar\Lambda_cA}_1}{q^2-m_A^2+i\,m_A\Gamma_A}\right)
 .
 \label{eq:VSAP}
 \en
The minus signs in front of the strong couplings
$h^{\Lambda_c\bar\Lambda_cV,A}_1$ come from the minus sign of the
$g^{\mu \nu}$ part of the $X_{c\bar c}$ propagator in
Eq.~(\ref{eq:decay amp}).

We take
$f_V\cdot h^{\Lambda_c\bar\Lambda_cV}_1=f_A\cdot
h^{\Lambda_c\bar\Lambda_cA}_1=4$~GeV and show in Fig.~\ref{fig:BR
thru resonance} the plots of branching ratios for each kind of
resonance as functions of both the mass and the width of the
resonance. The branching fractions depend on $(f_{X_{c\bar
c}}\cdot h^{\Lambda_c\bar\Lambda_cV}_1)^2$ and $(f_{X_{c\bar
c}}\cdot h^{\Lambda_c\bar\Lambda_cA}_1)^2$ proportionally.
\begin{figure}
\centerline{\epsfig{file=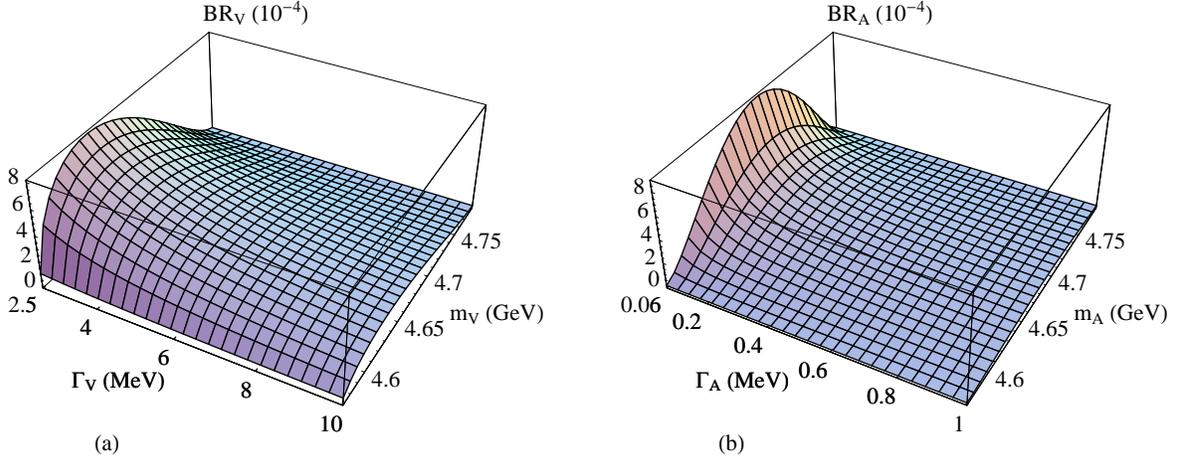,width=\textwidth}}
\caption{\label{fig:BR thru resonance}Branching fractions of
$\B_{X_{c\bar c}}(B^-\to \Lambda_c^+\bar\Lambda_c^-K^-)$ as a
function of the mass and the decay width of the intermediate
resonance $X_{c\bar c}$ when it is (a) a vector~($X_{c\bar c}=V$)
and (b) an axial-vector~($X_{c\bar c}=A$) particle with $f_V\cdot
h^{\Lambda_c\bar\Lambda_cV}_1=f_A\cdot
h^{\Lambda_c\bar\Lambda_cA}_1=4$~GeV.}
\end{figure}
We notice that the width of the axial-vector resonance $\Gamma_A$
is more than an order of magnitude smaller than $\Gamma_V$ when
both $\mathcal{B}_V$ and $\mathcal{B}_A$ are around the
experimental value $\mathcal{B}\sim7\times10^{-4}$. This is due
mainly to the smallness of $\mathcal{B}_A$ which suffers from the
destructive interference between {\it comparable} $\mathcal{M}_A$
and $\mathcal{M}_P$ contributions in the decay rate.
Fig.~\ref{fig:axial rate components} shows the decay rates from
$\mathcal{M}_A$, $\mathcal{M}_P$ and from
$\mathrm{Re}\mathcal{M}_A\mathcal{M}^*_P$ alone without taking
into account the resonance effect. One can see that the
$\mathcal{M}_A$ and $\mathcal{M}_P$ contributions are comparable
while the interference term
$\mathrm{Re}\mathcal{M}_A\mathcal{M}_P^*$ gives almost twice the
negative contribution of either $\mathcal{M}_A$ or
$\mathcal{M}_P$, resulting in a large cancellation with
$\mathcal{M}_A$ and $\mathcal{M}_P$.
\begin{figure}
\centerline{\epsfig{file=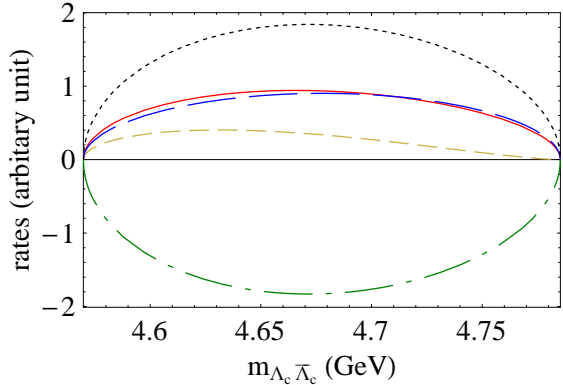,width=3.3in}}
\caption{\label{fig:axial rate components}Decay rates contributed
from $\mathcal{M}_V$~(short-dash), $\mathcal{M}_A$~(solid),
$\mathcal{M}_P$~(long-dash), sum of the previous two~(dotted) and
from $2\mathrm{Re}\mathcal{M}_A\mathcal{M}^*_P$~(dot-dashed) alone
without taking into account the resonance effect, i.e.
$(q^2-m_A^2+i\,m_A\Gamma_A)^{-1}$ is replaced by any constant in
each term.}
\end{figure}
Therefore, in order to counteract this cancellation, one needs a
smaller width of the resonance such that
$|1/(q^2-m_A^2+i\,m_A\Gamma_A)|^{-2}$ becomes more singular in the
allowed range of $q^2$.
There is, however, no such interference found in $\mathcal{B}_V$ as
$\mathcal{M}_V$ stands alone in the vector-induced decay amplitude
after taking $h^{\Lambda_c\bar\Lambda_cV}_2=0$ in
Eq.~(\ref{eq:vLcLc}).

Therefore, the above analysis seems to imply the existence of a
narrow hidden charm bound state with a mass of order $4.6\sim 4.7$
GeV that couples strongly with the charmed baryon pair. Recall
that many new charmonium-like resonances with masses around 4 GeV
starting with $X(3872)$ \cite{X3872} and so far ending with
$Y(4260)$ \cite{Y4260} have been recently observed by BaBar and
Belle. These charmonium-like states are above the $D\ov D$
threshold but below the two-charmed-baryon threshold. The new
state we have put forward is just marginally above the
$\Lambda_c\bar\Lambda_c$ threshold. In principle, this new state
can be searched for in $B$ decays and in $p\bar p$ collisions by
studying the mass spectrum of $D^{(*)}\bar D^{(*)}$ or
$\Lambda_c\bar\Lambda_c$.


%
 \section{Two-body decays}
\begin{figure}
\centerline{\epsfig{file=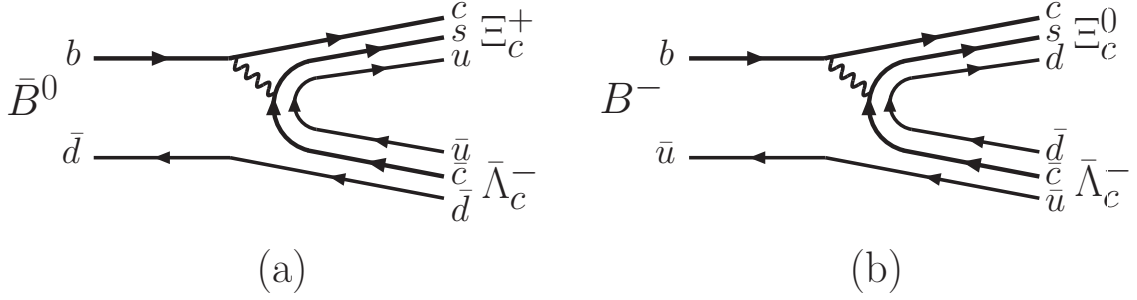,width=6in}}
\caption{\label{fig:B2CC} (a) $\ov B^0\to\Xi_c^+\bar\Lambda_c^-$
and (b) $B^-\to\Xi_c^0\bar\Lambda_c^-$ as proceeding through
internal $W$-emission diagrams.}
\end{figure}

The two-body doubly charmed baryonic $B$ decays
$B^-\to\Xi_c^0\bar\Lambda_c^-$ and $\ov
B^0\to\Xi_c^+\bar\Lambda_c^-$ receive contributions from the
internal $W$-emission (see Fig.~\ref{fig:B2CC}) and weak
annihilation. The latter contribution can be safely neglected as
it is not only quark-mixing but also helicity suppressed. It
should be stressed that, in contrast to the internal $W$-emission
in mesonic $B$ decays, internal $W$-emission in baryonic $B$ decay
is not necessarily color suppressed. This is because the baryon
wave function is totally antisymmetric in color indices. One can
see from Fig. \ref{fig:B2CC} that there is no color suppression
for the meson production. In the effective Hamiltonian approach,
the relevant weak Hamiltonian is
 \be \label{eq:effH}
 {\cal H}_{\rm eff}={G_F\over\sqrt{2}} V_{cb}V_{us}^*(c_1O_1+c_2O_2)
\to {G_F\over\sqrt{2}} V_{cb}V_{us}^*(c_1-c_2)O_1,
 \en
where $O_1=(\bar cb)(\bar sc)$ and $O_2=(\bar cc)(\bar sb)$. In
the above equation, we have used the fact that the operator
$O_1-O_2$ is antisymmetric in color indices (more precisely, it is
a color anti-triplet). Therefore, the Wilson coefficient for the
tree-dominated internal $W$-emission is $c_1-c_2$ rather than
$a_2=c_2+c_1/3$. This is indeed the case found in the pole model
calculation in \cite{CY02}.

Since the internal $W$-emission in Fig. \ref{fig:B2CC} is not
factorizable, it is difficult to evaluate its amplitude directly.
Pole model has been applied in \cite{CY02} to compute $B\to
\B_1\bar \B_2$. However, the strong coupling involved in this
model is unknown and hence it has to be fixed from other
processes, e.g. the 3-body baryonic $B$ decays. Since the CKM
angles for $\ov B^0\to\Xi_c^+\bar\Lambda_c^-$ and $\ov B^0\to
\Lambda_c^+\bar p$ are the same in magnitude (but opposite in
sign), the pole model does not explicitly explain why the former
has a rate much larger than the latter. In particular, the
dynamical suppression of $\Lambda_c^+\bar p$ relative to
$\Xi_c^+\bar\Lambda_c^-$ is not clearly manifested in the pole
model calculation. In order to understand why
$\Xi_c^+\bar\Lambda_c^-\gg \Lambda_c^+\bar p$, let us re-examine
Fig. \ref{fig:B2CC}.

There are several  possibilities for the quark-antiquark pair
creation in Fig. \ref{fig:B2CC}. In one case, $q\bar q$ is picked
up from the vacuum via the soft nonperturbative interactions so
that it carries the vacuum quantum numbers $^3P_0$. It is also
possible that the quark pair is created perturbatively via one
gluon exchange with one-gluon quantum numbers $^3S_1$. It is not
clear which mechanism, the $^3P_0$ or $^3S_1$ model, dominates the
2-body baryonic $B$ decays, though in practice the $^3P_0$ model
is simpler. Since the energy release is relatively small in
charmful baryonic $B$ decay, the $^3P_0$ model seems to be more
relevant. In the present work, we also consider the possibility
that the $q\bar q$ pair is produced via a light meson exchange.
The $q\bar q$ pair created from soft nonperturbative interactions
tends to be soft.  For an energetic proton produced in 2-body $B$
decays, the momentum fraction carried by its quark is large, $\sim
{\cal O}(1)$, while for an energetic charmed baryon, its momentum
is carried mostly by the charmed quark. As a consequence, the
doubly charmed baryon state such as $\Xi_c\bar\Lambda_c$ has a
configuration more favorable than $\Lambda_c\bar p$.

In order to evaluate Fig. \ref{fig:B2CC} for the decay $\ov
B\to\Xi_c\bar\Lambda_c$, we need to know the distribution
amplitudes of the charmed baryon $\mathcal{B}_c$ and the $B$
meson. For the wave functions of $\mathcal{B}_c=\Xi^0_c$,
$\Lambda_c$, they have the forms~\cite{Akhoury,Li}
 \be
 \la \Xi^0_c(p)|\bar c^a_\alpha(z_1) \bar s_\beta^b(z_2) \bar d_\gamma^c(z_3)|0\ra
 &=&{\epsilon^{abc}\over 6}{f_{\Xi_c}\over4}
 \left[\bar u_{\Xi^0_c}(p) \right]_\alpha
 \left[C^{-1}\gamma_5(p\!\!\!/+m_{\Xi_c})\right]_{\gamma\beta}
 \Psi_{\Xi^0_c}(z_1,z_2,z_3),
 \non\\
 \la \overline \Lambda_c(p')|c^a_\alpha(z'_1)u_\beta^b(z'_2)d_\gamma^c(z'_3)|0\ra
 &=&{\epsilon^{abc}\over 6}{f_{\Lambda_c}\over4}
  \left[\bar v_{\Lambda_c}(p') \right]_\alpha
  \left[(p\!\!\!/'-m_{\Lambda_c})\gamma_5C\right]_{\beta\gamma}
  \Psi_{\Lambda_c}(z'_1,z'_2,z'_3),
  \label{eq:baryonwf}
 \en
where $c,\,q$ and $d$ are the quark fields, $a,b$, and $c$ the
color indices, $\alpha,\beta$ and $\gamma$ the spinor indices, $C$
the charge conjugation matrix, and $f_{\B_c}$  the decay constant.
Following \cite{Braun} we can write
 \be
 \Psi_{\Xi_c}(z_1,z_2,z_3)
  &=& \int [dx][d^2k_\bot]\,e^{i\Sigma k_i \cdot z_i}\Psi_{\Xi_c}(x_1,x_2,x_3,{\bf k}_{1\bot},{\bf k}_{2\bot},{\bf k}_{3\bot}),
 \non\\
 \Psi_{\Lambda_c}(z'_1,z'_2,z'_3)
 &=& \int [dx'][d^2k'_\bot]\,e^{i\Sigma k'_i \cdot z'_i}
 \Psi_{\Lambda_c}(x'_1,x'_2,x'_3,{\bf k'}_{1\bot},{\bf k'}_{2\bot},{\bf k'}_{3\bot}),
 \en
where $p^{(\prime)}=(p^{(\prime)+},p^{(\prime)-},0_\bot)$ is the
momentum of $\Xi^0_c~(\bar\Lambda_c^-)$ and $k^{(\prime)}_1$,
$k^{(\prime)}_2$, $k^{(\prime)}_3$ the momenta of the constituent
quarks of the baryons, which are taken to be~\cite{Li,Li2000}
\be%
 &&k_1=(x_1p^+,p^-,\mathbf{k}_{1\bot})\,,\quad
    k^\prime_1=(p^{\prime+}, x^\prime_1p^{\prime-},\mathbf{k}^\prime_{1\bot})\,,
    \non\\
 &&k_{2(3)}=(x_{2(3)}p^+,0,\mathbf{k}_{2(3)\bot})\,,\quad
    k^\prime_{2(3)}=(0,x^\prime_{2(3)}p^{\prime-},\mathbf{k}^\prime_{2(3)\bot})
    \label{eq:kinamatics}
    \\
  &&[dx^{(\prime)}]  =  dx^{(\prime)}_1dx^{(\prime)}_2dx^{(\prime)}_3
  \delta\left(1-\sum_{i=1}^3x^{(\prime)}_i\right), \non \\
  && [d^2k^{(\prime)}_\bot]= d^2k^{(\prime)}_{1\bot}d^2k^{(\prime)}_{2\bot}d^2k^{(\prime)}_{3\bot}\delta^2
  \left({\bf k}^{(\prime)}_{1\bot}+{\bf k}^{(\prime)}_{2\bot}+{\bf k}^{(\prime)}_{3\bot}\right),
 \en
with $x_i$ being the momentum fractions associated with the
baryon, and ${\bf k}_{i\bot}$ the corresponding transverse
momenta.
Note that for simplicity, light quark masses are neglected in
Eq.~(\ref{eq:kinamatics}).
The $B$ meson wave function is expressed as
 \be \label{eq:Bwf}
 \la 0|\bar q^b_\alpha(z_1)b^a_\beta(z_2)|B(p)\ra=-i{\delta^{ab}\over3}{f_B\over 4}\left[(p\!\!\!/
+m_B)\gamma_5\right]_{\beta\alpha}\int_0^1d\xi\,e^{-i\Sigma
p_i\cdot z_i}\Phi_B(\xi)\,,
 \en
with $p_2=p_b=((1-\xi)p_B^+,(1-\xi)p_B^-,{\bf 0}_\bot)$,
$p_1=p_l=(\xi p_B^+,\xi p_B^-,{\bf 0}_\bot)$ and
$p_B^+=p_B^-=m_B$.

%
%
%

\begin{figure}
\centerline{\epsfig{file=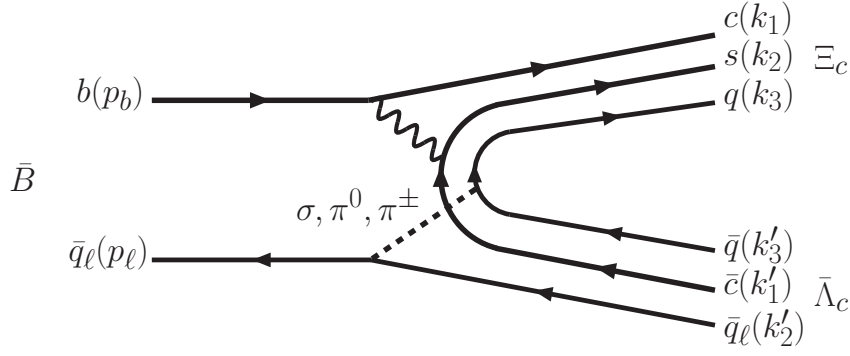,width=4.5in}}
\caption{\label{fig:sigma}Soft $q_{(\ell)}\bar q_{(\ell)}~[=u\bar
u~(d\bar d),\,d\bar d~(u\bar u)$ and $d\bar u~(u\bar d)]$ as
produced through the $\sigma,\,\pi^0$ and $\pi^\pm$ meson exchanges.
Inside the parentheses are the momenta that the constituent quarks
are carrying. Note the arrows on the quark lines do not represent
the momentum flows.}
\end{figure}

The $B^-\to\Xi_c^0\bar\Lambda_c^-$ decay amplitude now consists of
three parts corresponding to the exchange of the $\sigma$, $\pi^0$
and $\pi^-$ between the soft $q\bar q$ quark pair and the
spectator as shown in Fig.~\ref{fig:sigma},
\[
A(B^-\to\Xi_c^0\bar\Lambda_c^-)=A_\sigma+A_{\pi^0}+A_{\pi^-}\,,
\]
where, for instance, in the case that the decay proceeds through the
exchange of $\sigma$ or $\pi^0$
 \be
 A_{\sigma(\pi^0)}
 &=& {G_F\over\sqrt{2}} V_{cb}V_{us}^*(c_1-c_2)\int
 dz\,dz'\,e^{i(k_3+k'_3)\cdot z}e^{i(k'_2-p_\ell)\cdot z'}
 g_{\sigma(\pi) qq}^2\,D_F(z-z') \non \\
 && \quad\times\, \la \Xi_c^0|\bar c^a_\alpha(0)\bar s^b_\beta(0)\bar d^c_\gamma(z)|0\ra
 \la \bar\Lambda_c| c^b_\delta(0)u^d_{\eta^\prime}(z')d^c_{\gamma^\prime}(z)|0\ra
 \la 0|\bar  u^d_\eta(z')b^a_\rho(0)|B^-\ra \non \\
 && \quad \times\, \left[\gamma_\mu(1-\gamma_5)\right]_{\alpha\rho}\left[\gamma^\mu
 (1-\gamma_5)\right]_{\beta\delta}\Gamma_{\gamma\gamma^\prime}\Gamma_{\eta\eta^\prime}
 \label{eq:sigmapi0amp} \\
 &=& {G_F\over\sqrt{2}} V_{cb}V_{us}^*{f_Bf_{\Xi_c}f_{\Lambda_c}\over 64}
 {(c_1-c_2)\over 18}\int
 d\xi\int[dx][dx'][d^2k_\bot][d^2k'_\bot](2\pi)^4 \non \\
 && \times\,
 \delta^4(p_\ell-k'_2-k'_3-k_3)\,\Psi_{\Xi_c}(x,{\bf k}_\bot)\,\Psi_{\Lambda_c}(x',{\bf k}'_\bot)
 \Phi_B(\xi)\,{g_{\sigma(\pi) qq}^2\over p_{\sigma(\pi)}^2-m_{\sigma(\pi)}^2+i\,m_{\sigma(\pi)}\,\Gamma_{\sigma(\pi)}}\non \\
 &&\times\bar
 u_{\Xi_c}\left[a_{\sigma(\pi)}+b_{\sigma(\pi)}\gamma_5\right]v_{\Lambda_c}
  \label{eq:BtoCC}
 \en
with $\Gamma_{\eta\eta^\prime}=\Gamma_{\gamma\gamma^\prime}=1$ for
$\sigma$ and
$\Gamma_{\eta\eta^\prime}=-\Gamma_{\gamma\gamma^\prime}=i\,\gamma_5$
for $\pi^0$ in Eq.~(\ref{eq:sigmapi0amp}). Note that the factor of
$1/18$ in Eq.~(\ref{eq:BtoCC}) is the color factor. For the
$\pi^\pm$ exchange, the terms $g_{\pi qq}$,
$\Gamma_{\gamma\gamma^\prime}$ and $\la \bar\Lambda_c|
c^b_\delta(0)u^d_{\eta^\prime}(z')d^c_{\gamma^\prime}(z)|0\ra$ in
$A_{\pi^0}$ are replaced by $\sqrt2 g_{\pi qq}$,
$-\Gamma_{\gamma\gamma^\prime}$ and $\la \bar\Lambda_c|
c^b_\delta(0)d^d_{\eta^\prime}(z')u^c_{\gamma^\prime}(z)|0\ra=-\la
\bar\Lambda_c|
c^b_\delta(0)u^d_{\eta^\prime}(z)d^c_{\gamma^\prime}(z')|0\ra$,
respectively, where Eq.~(\ref{eq:baryonwf}) has been used. Due to
the symmetry property of the $\Lambda_c$ wave function given in
Eq.~(\ref{eq:Psi}), the $\pi^\pm$ contribution is the same as the
$\pi^0$ one except for an enhancement factor of $(\sqrt{2})^2$
arising from isospin.
Due to the tiny mass difference between $\pi^0$ and $\pi^-$ and
the extremely narrow widths of these two particles, we have
$A_{\pi^-}/A_{\pi^0}=2$ to a very good precision.
The momentum labels of the quarks in $\Xi_c$ and $\bar\Lambda_c$
are depicted in Fig. \ref{fig:sigma}, $p_b$ and $p_\ell$ are the
momenta of the $b$ quark and the light spectator quark of the $B$
meson defined after Eq~(\ref{eq:Bwf}), respectively,
$g_{\sigma(\pi) qq}$ is the coupling of the $\sigma~(\pi)$ meson
with the $q\bar q$ pair, and
 \be
 a_\sigma=-4m_{\Xi_c}\left[(m_B+m_{\Lambda_c})^2-m_{\Xi_c}^2\right],
 \qquad
 b_\sigma=-4m_{\Xi_c}\left[(m_{\Lambda_c}+m_{\Xi_c})^2-m_B^2\right]\,,
 \label{eq:a&b_sigma}
 \en
for $\sigma$ and
 \be \label{eq:a&b}
 a_\pi=4m_{\Xi_c}\left[m_B^2-(m_{\Xi_c}-m_{\Lambda_c})^2\right],
 \qquad
 b_\pi=4m_{\Xi_c}\left[m_{\Xi_c}^2-(m_B-m_{\Lambda_c})^2\right]\,,
 \label{eq:a&b_pi}
 \en
for both $\pi^0$ and $\pi^-$.

Our results are consistent with heavy quark effective theory. It
has been shown that in the heavy quark limit, the decay amplitude
can be expressed as $\bar u\left[A+B\gamma_5\right]v$
with~\cite{Mannel}
 \be \label{eq:A&B}
 A=2\sqrt{m_B}\,[\alpha(r_1-r_2)-\beta], \qquad\quad
 B=-2\sqrt{m_B}\,[\alpha(r_1+r_2)-\beta],
 \en
where $r_1=m_{\Xi_c}/m_B$, $r_2=m_{\Lambda_c}/m_B$ and $\alpha$,
$\beta$ are two unknown parameters. It is easily seen that,
Eqs. (\ref{eq:a&b_sigma}), (\ref{eq:a&b_pi}) and (\ref{eq:A&B})
agree with each other after setting
$\alpha=\alpha_\sigma+\alpha_\pi$, $\beta=\beta_\sigma+\beta_\pi$,
with $\alpha_\sigma=\beta_\sigma\propto 1+r_1+r_2$ for $\sigma$
and $\alpha_\pi=\beta_\pi\propto 1+r_1-r_2$ for both $\pi^0$ and
$\pi^-$, where the overall coefficients can be easily determined
from Eqs.~(\ref{eq:BtoCC}), (\ref{eq:a&b_sigma}) and
(\ref{eq:a&b_pi}).

To proceed with the numerical calculations, we need first deal
with the delta functions that impose constraints on the
integration limits as well as relations between integral
variables. We show in
the Appendix 
the decay amplitude and the integral variables as a result of the
delta function integrations.
The wave functions are adopted to be
 \be
 \Psi_{\mathcal{B}_c}(x_1,x_2,x_3)=\int[d^2k_\bot]\Psi_{\mathcal{B}_c}(x_i,{\bf
k}_{i\bot})
 =N_{\B_c}x_1x_2x_3\exp\left[-{\hat m_c^2\over
 2\beta^2x_1}-{\hat m_2^2\over 2\beta^2x_2}-{\hat m_3^2\over
 2\beta^2x_3}\right]\,,
 \en
with
\be%
\Psi_{\mathcal{B}_c}(x_i,{\bf
k}_{i\bot})=\frac{N_{\B_c}}{(2\pi\beta^2)^2} \exp\left[-{{\bf
k}^2_{1\bot}+\hat m_c^2\over
 2\beta^2x_1}-{{\bf k}^2_{2\bot}+\hat m_2^2\over 2\beta^2x_2}-{{\bf k}^2_{3\bot}+\hat m_3^2\over
 2\beta^2x_3}\right]\,,
 \label{eq:Psi}
\en
for the charmed baryon \cite{Schlumpf}  and
 \be
 \Phi_B(x)=N_Bx^2(1-x)^2\exp\left[-{1\over
 2}{x^2m_B^2\over\omega_B^2}\right]
 \en
for the $B$ meson \cite{Bwf}. The wave functions obey the
normalization
 \be
 \int [dx]\Psi(x_1,x_2,x_3)=1, \qquad\qquad \int^1_0 dx\Phi_B(x)=1.
 \en

The other input parameters are specified as follows. The decay
constants for the charmed baryons can be related to that of the
$\Lambda_b$ baryon via the relation \cite{Li2000}
 \be
 f_{\B_c}m_{\B_c}=f_{\Lambda_b}m_{\Lambda_b},
 \en
valid in the heavy quark limit. Using $f_{\Lambda_b}=2.71\times
10^{-3}$ GeV$^2$ obtained from a fit of the PQCD calculation for
$\Lambda_b\to\Lambda_c$ decays to
$\mathcal{B}(\Lambda_b\to\Lambda_cl\bar\nu)$ \cite{Li2000}, it is
found that $f_{\Lambda_c}=6.7\times 10^{-3}$ GeV$^2$ and
$f_{\Xi_c}=6.2\times 10^{-3}$ GeV$^2$, which are roughly $\sim
1.3-2.3$ times that of the results from QCD sum
rules~\cite{Colangelo:1995qp}.~\footnote{The $\Lambda_b$ decay
constant is found to be in the range $(2.0-3.5)\times
10^{-2}$~GeV$^3$ in QCD sum rules~\cite{Colangelo:1995qp}. It
differs from the $\Lambda_b$ decay constant in this work by a factor
of $\Lambda_b$ mass. After normalizing it to having the same
dimension as the decay constants in this work, the above range turns
out to be about $1.3-2.3$ times that of the decay constant from
PQCD.} In our calculations we shall employ the value of the
$\Lambda_b$ baryon decay constant which is in the middle of the
range 
that has the PQCD value as
the lower bound and the highest value from QCD sum rules as the
upper bound. The deviations of the lower and the upper bounds from
this central value are then taken as one of the theoretical errors
in our model.

For the $B$ meson, we use $f_B=0.2$ GeV. For the coupling
$g_{\sigma(\pi) qq}$, the linear sigma model leads to $g_{\sigma
NN}=\sqrt{2}\,m_N/f_\pi$ with $f_\pi=132$ MeV, and $g_{\pi
NN}=\sqrt{2}\,m_N g_A/f_\pi$ with $g_A\simeq1.25$ from the
Goldberger-Treiman relation. Hence, it is reasonable to take
$g_{\sigma(\pi) qq}=g_{\sigma(\pi) NN}/3$ in the constituent quark
model. For the $\sigma$ meson, we use $\Gamma_\sigma\approx
m_\sigma=600$ MeV \cite{PDG}. The constituent quark masses appearing
in the $\B_c$ wave function are taken to be $\hat m_u=\hat
m_d=0.33$~GeV and $\hat m_s=0.55$~GeV \cite{Schlumpf}, while $\hat
m_c=m_{{\mathcal B}_c}$ is employed.

The decay rate is given by
 \be
 \Gamma(B\to \B_1\ov \B_2)&=& {p_c\over 4\pi}\Bigg\{
 |A|^2\,{(m_B+m_1+m_2)^2p_c^2\over (E_1+m_1)(E_2+m_2)m_B^2}
 +|B|^2\,{[(E_1+m_1)(E_2+m_2)+p_c^2]^2\over
 (E_1+m_1)(E_2+m_2)m_B^2} \Bigg\},
 \en
where $p_c$ is the c.m. momentum, $E_i$ and $m_i$ are the energy and
mass of the baryon $\B_i$, respectively. The results of calculations
are summarized in Table \ref{tab:result}. The theoretical errors
come from the uncertainties in the parameters $\beta$, $\omega_b$,
which are taken to be $\beta=1.20\pm0.05$~GeV \footnote{As shown in
\cite{Schlumpf}, the parameter $\beta$ is of order 1 GeV for light
baryons. Just as the meson case \cite{CCH}, $\beta$ should become
larger for the heavy baryons.},
$\omega_b=0.40\pm0.05$~GeV, and the baryon decay constants. It is
clear that the pion exchange gives the dominant contribution owing
to its narrow width. The prediction is in agreement with experiment
for $\Xi_c^+\bar\Lambda_c^-$, but a bit small for
$\Xi_c^0\bar\Lambda_c^-$

\begin{table}[t]
\caption{Predictions on the branching ratios of
$B^-\to\Xi_c^0\bar\Lambda_c^-$ and $\ov
B^0\to\Xi_c^+\bar\Lambda_c^-$ decays. The first and second errors
come from the theoretical uncertainties in the parameters $\beta$
and $\omega_b$, respectively, which are taken to be
$\beta=1.20\pm0.05$~GeV and $\omega_b=0.40\pm0.05$~GeV, and the
third error from the baryon decay constants. Results shown in second
and third raws are from $\pi$ or $\sigma$ exchange alone,
respectively.}
 \label{tab:result}
\begin{ruledtabular}
\begin{tabular}{l r c |l  r c}
  Mode
  & Theory $(10^{-3})$
  & Expt $(10^{-3})$
  & Mode
  & Theory $(10^{-3})$
  & Expt $(10^{-3})$
  \\
\hline
  $\B(B^-\to\Xi_c^0\bar\Lambda_c^-)$
 & $2.2^{+0.6+5.1+6.1}_{-0.6-1.9-1.9}$
 & $\approx 4.8$
 & $\B(\ov B^0\to\Xi_c^+\bar\Lambda_c^-)$
 & $2.0^{+0.5+4.7+5.6}_{-0.6-1.7-1.7}$
 & $\approx 1.2$
 \\
 $ \B(B^-\to\Xi_c^0\bar\Lambda_c^-)_{\pi}$
 & $1.8^{+0.5+4.2+5.0}_{-0.5-1.6-1.6}$
 &
 & $\B(\ov B^0\to\Xi_c^+\bar\Lambda_c^-)_{\pi}$
 & $1.7^{+0.5+3.9+4.7}_{-0.5-1.4-1.5}$
 &
 \\
 $\B(B^-\to\Xi_c^0\bar\Lambda_c^-)_{\sigma}$
 &$0.2^{+0.0+0.4+0.6}_{-0.1-0.1-0.2}$
 &
 &$\B(\ov B^0\to\Xi_c^+\bar\Lambda_c^-)_{\sigma}$
 & $0.2^{+0.0+0.4+0.6}_{-0.0-0.1-0.2}$
 &
 \\
\end{tabular}
\end{ruledtabular}
\end{table}

The above calculation is not applicable to the two-body decay $\ov
B^0\to\Lambda_c^+\bar p$ with one charmed baryon in the final
state. This is because two hard gluons are needed to produce an
energetic antiproton: one hard gluon for kicking the spectator
quark of the $B$ meson to make it energetic and the other for
producing the hard $q\bar q$ pair. The pQCD calculation for this
decay will be much more involved (see e.g. \cite{Li} for pQCD
calculations of $\Lambda_b\to\Lambda J/\psi$) and is beyond the
scope of the present work. Nevertheless, it is expected that
$\Gamma(\ov B\to \B_c\bar N)\ll \Gamma(\ov
B\to\Xi_c\bar\Lambda_c)$ as the former is suppressed by order of
$\alpha_s^4$. This dynamical suppression effect for the
$\Lambda_c\bar p$ production relative to $\Xi_c\bar\Lambda_c$ has
been neglected in the previous studies based on QCD sum rules
\cite{Chernyak} and on the diquark model \cite{Ball}.

 \section{Conclusions}
In this work we have studied the two-body and three-body doubly
charmed baryonic $B$ decays, namely, $\ov B\to\Xi_c\bar\Lambda_c$
and $\ov B\to\Lambda_c\bar\Lambda_cK$, aiming to resolve the
puzzles associated with them. We point out that the suppression
from the $\Lambda_c^+\bar\Lambda_c^-K^-$ phase space is so strong
that $\Lambda_c^+\bar\Lambda_c^-$ pair is unlikely to be produced
dominantly through the direct three-body decay processes.
Nevertheless, the great suppression for the
$\Lambda_c^+\bar\Lambda_c^-K$ production can be alleviated
provided that there exists a narrow hidden charm bound state with
a mass near the $\Lambda_c\bar\Lambda_c$ threshold, of order
$4.6\sim 4.7$ GeV. This new state that couples strongly to the
charmed baryon pair can be searched for in $B$ decays and in
$p\bar p$ collisions by studying the mass spectrum of $D^{(*)}\bar
D^{(*)}$ or $\Lambda_c\bar\Lambda_c$.

The doubly charmful decay such as $\ov B\to\Xi_c\bar\Lambda_c$ has a
configuration more favorable than the singly charmful one such as
$\ov B^0\to\Lambda_c\bar p$ even though they have the same CKM
angles in magnitude. This is because no hard gluon is needed to
produce the energetic $\Xi_c\bar\Lambda_c$ pair in the former decay,
while two hard gluons are needed for the latter process. Therefore,
$\Lambda_c\bar p$ is suppressed relative to $\Xi_c\bar\Lambda_c$ due
to a dynamical suppression from ${\cal O}(\alpha_s^4)$. Assuming
that a soft $q\bar q$ quark pair  is produced through the $\sigma$
and $\pi$ meson exchanges  in the configuration for $\ov B\to
\Xi_c\bar\Lambda_c$, it is found that $\B(\ov
B\to\Xi_c\bar\Lambda_c)\sim 10^{-3}$.

\vskip 2.0cm \acknowledgments
 This research was supported in part
by the National Science Council of R.O.C. under Grant Nos.
NSC94-2112-M-001-016, NSC94-2112-M-001-023, NSC94-2811-M-001-050
and NSC94-2811-M-001-059.

\vskip 1.5 cm

\noindent {\it Note added:} After this paper was submitted for
publication, Belle published the updated version of \cite{3body}, in
which the spectrum of the $\Lambda_c\bar\Lambda_c$ pair in the
$\overline B\to\Lambda_c\bar\Lambda_cK$ decays is shown for the
first time. According to Belle's observation no new resonance with a
mass near the $\Lambda_c\bar\Lambda_c$ threshold was found (see Fig.
3 in version 2 of \cite{3body}). This implies the failure of naive
factorization for this decay mode and may hint at the importance of
nonfactorizable contributions such as final-state effects. For
example, the weak decay $B\to D^{(*)}\bar D_s^{(*)}$ followed by the
rescattering $D^{(*)}\bar D_s^{(*)}\to \Lambda_c\bar\Lambda_c
K$~\cite{Chen:2006fs} or the decay $B\to \Xi_c\bar\Lambda_c$
followed by $\Xi_c\bar\Lambda_c\to\Lambda_c\bar\Lambda_cK$ may
explain the large rate observed for $B\to\Lambda_c\bar\Lambda_cK$.

\appendix

\section{\label{sec:integration}}
After integrating out the delta functions in Eq.~(\ref{eq:BtoCC}),
we obtain, taking decay via $\sigma$ as an example,
\be \label{eq:noDelta} %
&& A_\sigma(B^-\to\Xi_c^0\bar\Lambda_c^-)\non\\
&& ={G_F\over\sqrt{2}}
V_{cb}V_{us}^*\,{f_Bf_{\Xi_c}f_{\Lambda_c}\over 1152}
 (c_1-c_2)(2\pi)^4
2 \int_0^{p^{\prime-}/p_B^-}
 d\xi\int_0^{1-\xi p_B^+/p^+}\frac{dx_2}{p^+}\int_0^
 {\xi p_B^-/p^{\prime-}}\frac{dx'_2}{p^{\prime-}}\non\\
&& \qquad\times(2\pi)\left(\frac{1}{2}\right)^3
\int^\infty_0d\mathbf{k}^2_{3\bot}\int^\infty_0d\mathbf{k}
^2_{2\bot}\int^{2\pi}_0d\theta_{23}\int^\infty_0d\mathbf{k}^{\prime2}_{3\bot}
 \int^{2\pi}_0d\theta_{33'}\non \\
&& \qquad\times
 \,\Psi_{\Xi_c}(x,{\bf k}_\bot)\,\Psi_{\Lambda_c}(x',{\bf k}'_\bot)
 \Phi_B(\xi)\,{\left(g_{\sigma qq}\right)^2\over p_\sigma^2-m_\sigma^2+i\,m_\sigma\Gamma_\sigma}
 \,\bar u_{\Xi_c}(a_\sigma+b_\sigma\gamma_5)v_{\Lambda_c}\,, 
 \en%
where $p^{(\prime)}=(p^{(\prime)+},p^{(\prime)-},0_\bot)$ is the
momentum of $\Xi^0_c~(\bar\Lambda_c^-)$ and $k^{(\prime)}_1$,
$k^{(\prime)}_2$, $k^{(\prime)}_3$ the momenta of the constituent
quarks of the baryons, which are taken to be
\be%
k_1=(x_1p^+,p^-,\mathbf{k}_{1\bot})\,,&&k^\prime_1=(p^{\prime+},
x^\prime_1p^{\prime-},\mathbf{k}^\prime_{1\bot})\,,\non\\
k_{2(3)}=(x_{2(3)}p^+,0,\mathbf{k}_{2(3)\bot})\,,&&
k^\prime_{2(3)}=(0,x^\prime_{2(3)}p^{\prime-},\mathbf{k}^\prime_{2(3)\bot})\,.
\en
The integrations of the delta functions yield the following
relations
\be%
&&x_1^{(\prime)}=1-x_2^{(\prime)}-x_3^{(\prime)}\,,\quad
x_3=\frac{\xi p^+_B}{p^+}\,,\quad x_3^\prime=\frac{\xi
p^-_B}{p^{\prime-}}-x_2^\prime\,,\non\\
&&\mathbf{k}^{(\prime)}_{1\bot}=-\left(\mathbf{k}^
{(\prime)}_{2\bot}+\mathbf{k}^{(\prime)}_{3\bot}\right)\,,
\quad
\mathbf{k}^\prime_{2\bot}=-\left(\mathbf{k}_{3\bot}+
\mathbf{k}^\prime_{3\bot}\right)\,\Rightarrow
\,\mathbf{k}^\prime_{1\bot}=\mathbf{k}_{3\bot}\,, \en
and the limits of integrations as shown in Eq.~(\ref{eq:noDelta}).
Thus,
\be%
p^2_\sigma=(k_3+k^\prime_3)^2&=&\left(\frac{\xi
p^+_B}{p^+}\right)\left(\frac{\xi
p^-_B}{p^{\prime-}}-x_2^\prime\right)p^+p^--\left(\mathbf
{k}^2_{3\bot}+\mathbf{k}^{\prime2}_{3\bot}
+2|\mathbf{k}_{3\bot}||\mathbf{k}^\prime_{3\bot}|\cos\theta_{33^\prime}\right)\,,\non\\
\mathbf{k}^2_{1\bot}&=&\mathbf{k}^2_{2\bot}+\mathbf{k}
^2_{3\bot}+2|\mathbf{k}_{2\bot}||\mathbf{k}_{3\bot}|\cos\theta_{23}\,,\non\\
\mathbf{k}^{\prime2}_{2\bot}&=&\mathbf{k}^2_{3\bot}+\mathbf{k}^{\prime2}_{3\bot}
+2|\mathbf{k}_{3\bot}||\mathbf{k}^\prime_{3\bot}|\cos\theta_{33^\prime}\,,
\en
where $\theta_{23}$ and $\theta_{33^\prime}$ are the angles of
$\mathbf{k}_{2\bot}$ and of $\mathbf{k}^\prime_{3\bot}$ as
measured against $\mathbf{k}_{3\bot}$, respectively.
Note that the integration ranges of $x_2$ and $x'_2$ are
constrained by the delta function in Eq.~(\ref{eq:BtoCC}), while
the integration range of $\xi$ is restricted by $(1-\xi
p_B^+/p^+)\geq0$ and $\xi p_B^-/p^{\prime-}\leq1$.


\end{document}